\documentclass[%
 aip,
 amsmath,amssymb,
preprint,%
]{revtex4-1}

\usepackage{graphicx}
\usepackage{dcolumn}
\usepackage{bm}

\usepackage{lipsum} 
\usepackage{hyperref}
\hypersetup{
    colorlinks=true,
    linkcolor=blue,
    citecolor=blue,
    urlcolor=blue,
    pdftitle={Overleaf Example},
    pdfpagemode=FullScreen,
    }

\newcommand{\expect}[1]{\mathbb{E}#1}
\newcommand{\variance}[1]{\mathbb{V}#1}

\newcommand{\identity}{\mathbb{I}}

\newcommand{\ceil}[1]{\left \lceil #1 \right \rceil}
\newcommand{\floor}[1]{\left \lfloor #1 \right \rfloor}

\usepackage{xcolor}

\usepackage[utf8]{inputenc}
\usepackage[T1]{fontenc}
\usepackage{mathptmx}
\usepackage{etoolbox}

\usepackage{tikz}
\usetikzlibrary{quantikz2}

\makeatletter
\def\@email#1#2{%
 \endgroup
 \patchcmd{\titleblock@produce}
  {\frontmatter@RRAPformat}
  {\frontmatter@RRAPformat{\produce@RRAP{*#1\href{mailto:#2}{#2}}}\frontmatter@RRAPformat}
  {}{}
}%
\makeatother
\begin{document}


\title{Higher order tensor factorizations for block encoding vibrational and vibronic Hamiltonians}
\author{Hirsh Kamakari}
\email{hirsh@beit.tech}
\affiliation{BEIT Canada Inc., 101 College St, Toronto, Canada}
\altaffiliation{https://www.beit.tech}

\author{Emil Zak}
\affiliation{BEIT sp. z o o., Mogilska 43, 31-545 Krak\'ow, Poland}
\altaffiliation{https://www.beit.tech}


\date{\today}

\begin{abstract}
Fault tolerant quantum simulation via the phase estimation algorithm and qubitization has a T-gate count that scales proportionally to the 1-norm of the Hamiltonian, the cost of block encoding the Hamiltonian, and inversely proportionally to the desired accuracy. Tensor factorization methods have been successfully used to reduce T-gate counts in the ground state electronic structure problem. Here we introduce the use of tensor factorization methods to reduce the T-gate count of quantum phase estimation. In particular, we show how Canonical Polyadic and Tucker decompositions of the tensors representing the vibrational and vibronic Hamiltonians can be utilized to rewrite the Hamiltonian in terms of linear combination of bosonic position operators representing nuclear vibrations. We demonstrate the use of these factorization methods on the water and monodeutered methane molecules.
\end{abstract}

\maketitle

\section{\label{sec:intro}Introduction}

Vibrational and vibronic dynamics play a fundamental role in understanding phenomena such as non-radiative relaxation processes, energy transfer, and photochemical processes as well as being key to understanding molecular spectroscopy data.  Understanding these processes is essential for developing higher efficiency solar cells \cite{long2017, panhans2020, desio2017}, molecular junctions \cite{erpenbeck2016, hartle2009, kushmerick2004}, cancer therapies \cite{ayala2019, ayala2024}, catalysis \cite{he2020, lather2019}, and molecular spectroscopy \cite{kong2021, mchale2017, barone2021}. 

While various classical computational chemistry tools have been developed for the simulation of vibrational and vibronic dynamics\cite{zobel2021}, as with the electronic structure problem, many solutions rely on semiclassical approximations or on resource intensive algorithms which scale poorly with the system size~\cite{Christiansen2004,Xie2019-vn,MendiveTapia2012,Giri2011,Curchod2018,Avila2011}. As a result, several digital quantum algorithms have been proposed for the simulation of vibrational dynamics \cite{malpathak2024, mcardle2019, jahangiri2020, sawaya2020, trenev2025} and vibronic dynamics \cite{ollitrault2020, motlagh2024a}. Analog quantum simulation of vibrational and vibronic coupling has also been investigated \cite{macdonell2021, kang2024}. Quantum simulation in first and second quantization has been explored as a means to simulate vibronic coupling as the electrons and nuclei can be treated on equal footing in a pre-Born-Oppenheimer approach \cite{su2021, mukhopadhyay2024, ha2024}.

Despite the moderate amount of research done in the area of quantum simulation of vibrational and vibronic interactions, little attention has been focused on applying the state of the art quantum simulation algorithms to the vibronic case. Qubitization\cite{low2017, low2019} and various Hamiltonian factorization methods have been successfully used to efficiently embed electronic structure problems in a quantum computer. It has been shown that by considering low rank approximations to the electronic Hamiltonian, the quantum computing resources required to solve the ground state problem can be reduced \cite{motta2021, berry2019}. Further savings can be achieved by the use of double-factorized and tensor hypercontracted Hamiltonians \cite{burg2021, lee2021, loaiza2024, oumarou2024}. Factorization can be combined with other techniques such as filter diagonalization\cite{cohn2021}, spectrum amplification\cite{low2025} and the use of symmetry considerations\cite{rocca2024, deka2024, caesura2025} to achieve further savings.

In this work, we introduce the use of tensor factorization methods to allow for efficient block encodings of ab intio vibrational and vibronic Hamiltonians. In particular, we show how Tucker and Canonical Polyadic (CP) decompositions can be used to reduce the gate counts for phase estimation by reducing the 1-norm of the Hamiltonians at the cost of increased gate counts for the block encoding quantum circuit.

In Section \ref{sec:methodology}, we introduce the methodology of using CP and Tucker decompositions to represent higher order tensors appearing in vibrational and vibronic Hamiltonians. In Section \ref{sec:results} we present fault-tolerant quantum computing resource counts for a model of the water and monodeutered methane molecule and demonstrate the reduction in resources that can be achieved. Finally, in Section \ref{sec:discussion} we discuss our results, comparing the different factorization methods as well as the advantages and drawbacks of our approach to previous methods. 
 
\section{\label{sec:methodology}Methodology}

\subsection{\label{subsec:Hamiltonians}Vibrational and vibronic Hamiltonians}
A general Hamiltonian describing electronic, vibrational, and vibronic couplings can be written in second quantization in terms of the fermionic creation and annihilation operators $c_{i\sigma}^\dagger,c_{i\sigma}$ and the bosonic creation and annihilation operators $b_\alpha^\dagger,b_\alpha$ satisfying appropriate anti-commutation and commutation relations, respectively. The fermionic operators describe the electronic excitations of the molecule and the bosonic operators describe the nuclear motion state excitations with amplitudes determined by the molecular potential energy surface (PES). In the following, we label electronic degrees of freedom by the letters $i,j$ (orbitals), $\sigma$ (spin) and bosonic degrees of freedom by the letters $\alpha,\beta$.  A Hamiltonian describing $N$ electrons and $M$ vibrational modes can be written as~\cite{Csszr2011,Hirata2014,Baiardi2017,Sibaev2020,trenev2025}
\begin{equation}
H=H_{\mathrm{el}} + H_{\mathrm{v}}+H_{\mathrm{vc}} = H_{\mathrm{el}} + H_{\mathrm{vvc}},
\label{eq:Hamiltonian}
\end{equation}
where $H_{\mathrm{el}}$ is the electronic Hamiltonian
\begin{equation} \label{eqn:Hamiltonian_form_first}
	H_{\mathrm{el}} = \sum_{i, j, \sigma} h_{ij} c_{i \sigma}^{\dag} c_{j \sigma}
	+ \sum_{\substack{i, j,k,l \\ \sigma, \tau }} g_{ijkl} c_{i \sigma}^{\dag} c_{j \sigma} c_{k \tau}^{\dag} c_{l \tau},
\end{equation}
$H_{\mathrm{v}}$ is the vibrational Hamiltonian
\begin{equation}
H_{\mathrm{v}} = H_{\mathrm{harmonic}} + \sum_{k=3}^{L_{\mathrm{v}}}\left(\sum_{\alpha_1\cdots \alpha_k}E_{\alpha_1\dots \alpha_k}q_{\alpha_1}\cdots q_{\alpha_k}\right),
\label{eq:high_order_vibrational}
\end{equation}
and $H_{\mathrm{vc}}$ is the vibronic coupling interaction
\begin{equation}
    H_{\mathrm{vc}} = \sum_{k=1}^{L_{\mathrm{vc}}}\sum_{\substack{\alpha_1\cdots\alpha_k \\ i,j,\sigma}}E_{\alpha_1\cdots\alpha_kij\sigma }q_{\alpha_1}\cdots q_{\alpha_k} c_{i\sigma}^\dagger c_{j\sigma}.
    \label{eq:higher_order_vibronic}
\end{equation}
where the bosonic displacement operator is defined as $q_\alpha=(b_\alpha + b_\alpha^\dagger)/\sqrt{2}$.
In Eq.~\eqref{eq:high_order_vibrational}, $H_{\mathrm{harmonic}}=\sum_\alpha\omega_\alpha b_\alpha^\dagger b_\alpha$ is the harmonic oscillator vibrational Hamiltonian and $L_{\mathrm{v}}$  is the truncation order for vibrational interactions. Similarly in Eq.~\eqref{eq:higher_order_vibronic}, $L_{\mathrm{vc}}$ is the order of the truncation of the vibronic coupling. We note here that the tensors $E_{\alpha_1\dots \alpha_k}$ are totally symmetric and that the tensors $E_{\alpha_1\cdots\alpha_k\sigma ij}$ are symmetric in the $\alpha$ indices and $i,j$ indices separately. In Eqs.~\eqref{eq:high_order_vibrational} and \eqref{eq:higher_order_vibronic}, $\alpha_j=0,1,2,...,M-1$ label vibrational modes. 

In order to simulate a molecular system described by the Hamiltonian of Eq.~\eqref{eq:Hamiltonian}, the bosonic and fermionic operators need to be encoded into qubit operators. The fermionic operators can be encoded using the Jordan-Wigner transformation \cite{jordan1928, somma2002} and the bosonic operators can be encoded using a unary encoding of the Fock space \cite{sawaya2020, tudorovskaya2024}. 

The most efficient quantum algorithms for implementing time evolution generated by the Hamiltonian $H$ in Eq.~\eqref{eq:Hamiltonian} for estimating the low energy vibrational spectrum via quantum signal processing requires a block encoding of $H$ \cite{low2017, low2019, gilyen2019, motlagh2024b}. For estimating the low energy vibrational spectrum via phase estimation in particular, the circuit depth and gate count scales proportionally to $\lambda B$, the 1-norm $\lambda$ of the Hamiltonian and the cost $B$ of block encoding, respectively \cite{babbush2018}. The 1-norm of the vibronic Hamiltonian in Eq.~\eqref{eq:Hamiltonian} can be bounded by 
\begin{align}
    \lambda &= |H_{\mathrm{vvc}}|_1\leq |H_{\mathrm{v}}|_1+|H_{\mathrm{vc}}|_1\\
    &= \sum_\alpha|\omega_\alpha| +  \sum_{k=3}^{L_{\mathrm{v}}}\sum_{\alpha_1\cdots \alpha_k}|E_{\alpha_1\dots \alpha_k}| + \sum_{k=1}^{L_{\mathrm{vc}}}\sum_{\substack{\alpha_1\cdots\alpha_k \\ i,j,\sigma }}|E_{\alpha_1\cdots\alpha_kij\sigma }|.
    \label{eq:1norm}
\end{align}

\subsection{\label{subsec:mutlifactorization}Mutlifactorization of high order tensors}
Minimizing $\lambda$ in Eq.~\eqref{eq:1norm} is crucial for reducing the T-gate count in quantum phase estimation. In this section, we show how CP and Tucker decompositions of the vibrational and vibronic Hamiltonians can be used to reduce $\lambda$. 

As shown in Refs. \cite{comon2008, kolda2015}, a symmetric tensor $E_{\alpha_1\cdots \alpha_k}$ always has a CP decomposition 
\begin{equation}
    E_{\alpha_1\cdots \alpha_k}= \sum_{l=1}^{r_k}\Lambda_{kl}Q_{kl\alpha_1}\cdots Q_{kl\alpha_k}
    \label{eq:cp}
\end{equation}
where each $Q_{kl}$ is a unit vector and $r_k$ is the rank of the approximation. Contrary to the case for matrices, computing the minimal rank $r_k$ such that the approximation in Eq.~\eqref{eq:cp} is exact is NP-hard even in the case of symmetric tensors \cite{hillar2013}; however, various algorithms exist which can compute such an approximation for a given rank\cite{kolda2009,battaglino2018,erichson2020}.

Substituting the decomposition of Eq.~\eqref{eq:cp} into the vibrational Hamiltonian $H_\mathrm{v}$ yields 
\begin{equation}
    H_\mathrm{v} = H_{\mathrm{harmonic}} + \sum_{k=3}^{L_\mathrm{v}}\sum_{l=1}^{r_k}\Lambda^\mathrm{v}_{kl}(s_{kl})^k, \quad s_{kl} = \sum_{\alpha}Q^\mathrm{v}_{kl\alpha}q_{\alpha}.
    \label{eq:cp_vibrational_decomp}
\end{equation}
We substitute a similar decomposition into the vibronic Hamiltonian given in Eq.~\eqref{eq:higher_order_vibronic}. However, since the tensors describing the vibronic Hamiltonians are not totally symmetric and only symmetric in the $\alpha$ indices and electronic indices separately, we have a separate decomposition for each set of $(\sigma,i,j)$. The resulting factorized vibronic Hamiltonian is given by     
\begin{equation}
    H_{\mathrm{vc}} = \sum_{k=1}^{L_{\mathrm{vc}}}\sum_{\sigma, i, j}\sum_{l=1}^{r_k}\Lambda^\mathrm{vc}_{kli j\sigma }\left( s_{klij\sigma }\right)^k c_{i\sigma}^\dagger c_{j\sigma}, \quad s_{klij\sigma}=\sum_{\alpha}Q^{\mathrm{vc}}_{kl\alpha ij\sigma }q_{\alpha}.\label{eq:cp_vibronic_decomp}
\end{equation} Detailed derivations of both factorized Hamiltonians can be found in SI Section \ref{si_sec:cp_decomposition}.

The 1-norm of the resulting reduced Hamiltonian is given by 
\begin{equation}
    \lambda_{\mathrm{CP}} = \sum_\alpha|\omega_\alpha| +  \sum_{k=3}^{L_{\mathrm{v}}}\sum_{l=1}^{r_k}|\Lambda^\mathrm{v}_{k}| + \sum_{k=1}^{L_{\mathrm{vc}}}\sum_{\sigma, i, j}\sum_{l=1}^{r_k}|\Lambda^\mathrm{vc}_{kl i j\sigma}|.
    \label{eq:1norm_cp}
\end{equation}

We next show how to use Tucker decomposition\cite{lathauwer2000} to reduce the 1-norm of $H_\mathrm{v}$ and $H_\mathrm{vc}$. The Tucker decomposition of a symmetric tensor has the form
\begin{equation}
    E_{\alpha_1\cdots \alpha_k} = \sum_{\beta_1\cdots \beta_k}\Lambda_{\beta_1\cdots \beta_k}Q_{k\beta_1\alpha_1}\cdots Q_{k\beta_k\alpha_k},
    \label{eq:tucker_decomp}
\end{equation}
where each $Q_g$ is a unitary matrix. The advantage of the Tucker decomposition over the CP decomposition is that there exists efficient algorithms for the exact decomposition. 

If we substitute the decomposition from Eq.~\eqref{eq:tucker_decomp} into the vibrational Hamiltonian, we get the factorized Hamiltonian
\begin{equation}
    H_\mathrm{v} = H_{\mathrm{harmonic}} + \sum_{k=3}^{L_\mathrm{v}}\sum_{\beta_1\cdots \beta_k}\Lambda^\mathrm{v}_{\beta_1\cdots \beta_k}s_{k\beta_1}\cdots s_{k\beta_k}, \quad s_{k\beta} = \sum_{\alpha}Q_{k\beta\alpha}q_{\alpha} \label{eq:tucker_vibrational_decomp}
\end{equation}
Similarly for the factorized vibronic Hamiltonian, we have
\begin{equation}
    H_\mathrm{vc} = \sum_{k=1}^{L_\mathrm{vc}}\sum_{\beta_1\cdots \beta_k}\sum_{\sigma, i, j}\Lambda^\mathrm{vc}_{\beta_1\cdots \beta_k\sigma i j}s_{k\beta_1ij\sigma}\cdots s_{k\beta_kij\sigma}c_{i\sigma}^\dagger c_{j\sigma}, \quad s_{k\beta ij\sigma} = \sum_{\alpha}Q_{k\beta ij\sigma\alpha}q_{\alpha} \label{eq:tucker_vibronic_decomp}
\end{equation}
The 1-norm resulting from the Tucker decomposition is given by
\begin{equation}
    \lambda_{\mathrm{Tucker}} = \sum_\alpha|\omega_\alpha| +  \sum_{k=3}^{L_{\mathrm{v}}}\sum_{\beta_1\cdots\beta_k}|\Lambda^\mathrm{v}_{\beta_1\cdots\beta_k}| + \sum_{k=1}^{L_{\mathrm{vc}}}\sum_{\sigma, i, j}\sum_{\beta_1\cdots\beta_k}|\Lambda^\mathrm{vc}_{\beta_1\cdots\beta_ki j\sigma }|.
    \label{eq:1norm_cp}
\end{equation}

\subsection{\label{subsec:block_encoding}Block encoding of $H_{\mathrm{vvc}}$}
Although both CP and Tucker decompositions result in 1-norms lower than that of the original Hamiltonian, the block encoding can have an increased cost in both gate count and number of ancilla qubits that must be taken into account. The increase in cost comes from the fact that we must form linear combinations of the $q_\alpha$'s defined by the factors $Q_g$. In this work, we adopt the linear combination of unitaries (LCU)~\cite{childs2012} scheme for block-encoding the Hamiltonian, for which two quantum circuits must be provided: \textit{Prepare} and \textit{Select}. The \textit{Prepare} unitary circuit creates a multi-qubit state with amplitudes given by square roots of respective coefficients in the LCU representation of the Hamiltonian. Construction of \textit{Prepare} requires the Quantum Read Only Memory (QROM) oracle, for which we choose the SELECT-SWAP method from Ref.~\cite{low2024}.
For implementing the \textit{Select} unitaries we choose the unary iteration procedure\cite{gidney2018, babbush2018}, which selects $L$ elements using $4L-4$ T gates and $\ceil{\log L}$ ancilla qubits. For a detailed description of the block-encoding of the Hamiltonian and the associated resources, see SI Section \ref{si_sec:block_encoding}.
With block-encoding circuits for the $s_{k\beta ij\sigma}$ and $s_{k\beta}$ operators we construct block-encoding of the Hamiltonian $H_\mathrm{vvc}$ using techniques for addition and multiplication of block encodings~\cite{burg2021}. 
The block encoding procedure is done from the inner most terms of the decomposition of the Hamiltonians (cf. Eqs.~\eqref{eq:cp_vibronic_decomp},\eqref{eq:tucker_vibronic_decomp}) to the outer most terms. First, each individual $q_\alpha$ operator is block-encoded as it can be expressed as a linear combination of Pauli operators using a unary encoding of the truncated Fock space. Then, we use circuits for adding block encodings\cite{burg2021} to construct the linear combination of the $q_\alpha$'s that appear in equations \eqref{eq:cp_vibrational_decomp}, \eqref{eq:cp_vibronic_decomp}, \eqref{eq:tucker_vibrational_decomp}, and \eqref{eq:tucker_vibronic_decomp}.
Products of the $s$ operators can be block-encoded using standard circuits for the multiplication of block encodings\cite{burg2021}. Finally, the full Hamiltonian is block encoded by summing together all sub-block encodings, again using circuits for the addition of block encodings. We note that the circuits for adding linear combinations of Hamiltonians are equivalent to the divide-and-conquer or recursive block encodings discussed in ref.\cite{mukhopadhyay2024} and have identical gate counts.

\section{\label{sec:results}Results}
We first present asymptotic gate counts for block encoding the vibronic Hamiltonian given in Eq.~\eqref{eq:Hamiltonian} in terms of the Hamiltonian's parameters: the number of electronic orbitals $N$, the number of vibrational modes $M$, vibrational excitation cut-off $d$, the maximum number of coupled vibrational modes $L_\mathrm{v}$, and the maximum number of vibronically coupled modes $L_{\mathrm{vc}}$. The block encoding cost can be decomposed as $C = S+2P$, where $S$ is the cost of \textit{Select} and $P$ is the cost of \textit{Prepare}. We note that since we are using recursive block encodings, the inner block encodings have their own \textit{Select} and \textit{Prepare} costs. We combine all the \textit{Select} unitaries from all inner block encodings and all \textit{Prepare} unitaries from all inner block encodings, which does not affect the gate count. The asymptotic gate counts are derived in SI Section \ref{si_sec:asymptotic_gate_count} and here we present the final T-gate count upper bounds. 

The CP decomposition has a rank parameter $r$ defined in Eq.~\eqref{eq:cp_vibrational_decomp} for determining the level of approximation. For simplicity, we assume that the parameters $d$ and $r$ are constant for each order in the Taylor series expansion.  Using the unary iteration procedure\cite{babbush2018} for implementing the \textit{Select} operation and the SELECT-SWAP method\cite{low2024} for implementing the \textit{Prepare} operations, we have an asymptotic T-count of 
\begin{align}
O\left(rMd\left(L_\mathrm{v}^2+L_\mathrm{vc}^2N^2\right) + 
rL_\mathrm{v}^2\sqrt{Md}\log\frac{Md}{\epsilon'}\right.&+\sqrt{rL_\mathrm{v}}\log\frac{rL_\mathrm{v}}{\epsilon'}  \\ 
  +rL_\mathrm{vc}^2\sqrt{Md}\log\frac{Md}{\epsilon'}&+\left.N\sqrt{rL_\mathrm{vc}}\log\frac{rL_\mathrm{vc}N^2}{\epsilon'}\right)
      \label{eq:cost_cp}
\end{align}
where 
\begin{equation}
    \epsilon'=\frac{\epsilon_\mathrm{Prep}}{r(L_\mathrm{v}^2+L_\mathrm{vc}^2)}
\end{equation}
and $\epsilon_{\mathrm{Prep}}$ is the maximum allowable error in all state preparation operations. $\epsilon_{\mathrm{Prep}}$ is related to the 1-norm $\lambda$ and accuracy of the simulation  $\Delta E$  by $\epsilon_{\mathrm{Prep}}=(1/3\sqrt{2})\Delta E/\lambda$. For eigenenergy estimation to chemical accuracy, $\Delta E=1.6\mathrm{mH}$. See SI Section \ref{si_sec:error_analysis} for more details.

For the Tucker decomposition shown in Eqs.~\eqref{eq:tucker_vibrational_decomp} and \eqref{eq:tucker_vibronic_decomp}, we do not have a rank parameter, and instead use the full rank representation. This results in a total T-count of 
\begin{align}
    O\left(dL_\mathrm{v}M^{L_\mathrm{v}+1} + N^2dL_\mathrm{vc}M^{L_\mathrm{vc}+1} + L_\mathrm{v}M^{L_\mathrm{v}+1/2}\sqrt{d}\log\frac{Md}{\epsilon'} \right. &+ \\  +M^{L_\mathrm{v}/2}\log\frac{M^{L_\mathrm{v}}}{\epsilon'} + L_\mathrm{vc}M^{L_\mathrm{vc}+1/2}\sqrt{d}\log\frac{Md}{\epsilon'} &+ \left.NM^{L_\mathrm{vc}/2}\log\frac{M^{L_\mathrm{vc}}N^2}{\epsilon'}\right)
    \label{eq:cost_tucker}
\end{align}
where 
\begin{equation}
\epsilon'=\frac{\epsilon_\mathrm{Prep}}{L_\mathrm{v}M^{L_\mathrm{v}}+L_\mathrm{vc}M^{L_\mathrm{vc}}}.
\end{equation}
The main difference between the T-gate counts presented in Eqs.~\eqref{eq:cost_tucker} and \eqref{eq:cost_cp} is that the CP decomposition has a polynomial dependence on the order of approximations $L_\mathrm{v}$ and $L_\mathrm{vc}$, while the full rank Tucker decomposition has an exponential dependence on $L_\mathrm{v}$ and $L_\mathrm{vc}$. In the general case of dense Hamiltonians, phase estimation with the unfactored Hamiltonian also has an exponential T-count scaling in $L_\mathrm{v}$ and $L_\mathrm{vc}$. In this case, both CP and Tucker decompositions can lead to a decrease in T-count. 

\begin{figure}
    \centering
    \includegraphics[width=0.5\linewidth]{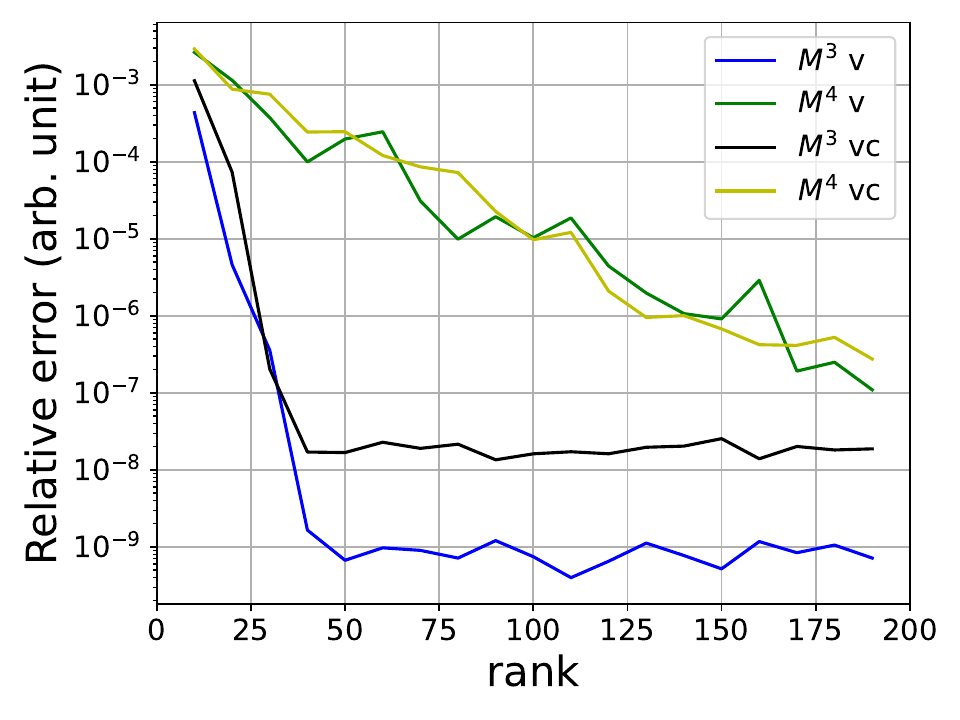}
    \caption{The relative error in the CP decomposition for the different tensors representing the H$_2$O molecule. $M^3$ indicates decomposition of a 3 index tensor (three-body coupling) while $M^4$ indicates decomposition of a 4 index tensor. v denotes the vibrational Hamiltonian and vc decnotes the vibronic Hamiltonian. }
    \label{fig:error_vs_rank_water}
\end{figure}

In the asymptotic gate counts for the CP decomposition given in Eq.~\eqref{eq:cost_cp}, we showed how the gate cost of a simulation depends on the rank $r$ of the decomposition. To determine the gate count only in terms of the parameters of the Hamiltonian, and not the rank, we need to choose the rank of the decomposition of each tensor $E_{\alpha_1\cdots\alpha_k}$ and $E_{\alpha_1\cdots\alpha_k\sigma ij}$ . The rank can be chosen to be the minimal rank such that the error in the approximation, $\epsilon_\mathrm{F}$, of each tensor is at most:
\begin{equation}
    \epsilon_{\mathrm{F}} \leq \frac{1}{3\sqrt{2}(L_\mathrm{v} -2 + N^2(L_\mathrm{vc} - 1))}\frac{\Delta E}{\lambda}.
\end{equation}
The prefactor in the denominator, $L_\mathrm{v} -2 + N^2(L_\mathrm{vc} - 1)$, is the number of tensors to be decomposed (see SI Section \ref{si_sec:error_analysis} for more details). 
Since there is no analytical relation between the ranks of the CP decompositions and the error in approximating the Hamiltonian, we show in Figure~\ref{fig:error_vs_rank_water} the empirical scaling of the approximation error with increasing rank for an example H$_2$O molecule Hamiltonian. The potential energy surface for H$_2$O was obtained from~\cite{huang2008}. For the purpose of demonstration, we use the vibrational Hamiltonian as a proxy for the vibronic Hamiltonian by randomly perturbing vibrational terms and applying an exponential damping factor dependent on the order of the tensor and the $i$ and $j$ indices. See SI Section \ref{si_sec:vibronic_hamiltonian} for more details. We define the relative error in approximating a tensor $E$ by a factorized tensor $E_\mathrm{F}$ as $\epsilon_\mathrm{F}=||E-E_\mathrm{F}||_2/||E||_2$ where $||\cdot||_2$ is the entry-wise 2-norm.  We present the 1-norm, T-counts and the relative costs of the CP and Tucker decompositions of the Hamiltonian for the phase estimation algorithm of the H$_2$O molecule and $\Delta E/\lambda=0.01$ in Table \ref{table:resources_water}.

\begin{table}[h]
\begin{center}
\begin{tabular}{ |c |c| c| c| c|c|c| } 
 \hline
  & $\lambda$ ($E_\mathrm{h}$) & Qubits & Select T count \quad & Prepare T count\quad & Total T count \quad & Relative cost \quad\\
 \hline\hline
 Unfactorized & 55.0 & 34 &$6.6\times 10 ^ 4$ & $1.1\times 10 ^ 4$ & $2.1\times10^9$ & 1\\ 
 Tucker decomposition & 91 & 32& $7.2\times10^3$ & $2.0\times10^4$ & $1.4\times10^9$ & 0.66\\ 
 CP decomposition & 52.1 & 34 & $6.7\times 10 ^ 4$ & $1.3\times 10 ^ 4$ & $2.2\times10^9$ & 1.05\\ 
 \hline
\end{tabular}
\caption{The 1-norms, T counts and relative cost of factorization methods to the unfactorized implementation of the CP and Tucker decompositions for the water molecule. For the CP decomposition, we used decomposition ranks corresponding to a total QPE error of 1\%. The 1-norm $\lambda$ is given in Hartrees ($E_\mathrm{h}$).}
\label{table:resources_water}
\end{center}
\end{table}

We finally present numerical results for the T-gate cost in the simulation of monodeutered methane (CH$_3$D) with 9-dimensional potential energy surface obtained from Ref.\cite{rey2014}. The 1-norm ($\lambda$), T-gate count for implementing \textit{Select} (S), the T-gate count for implementing \textit{Prepare} (P), and the total cost $C=\sqrt{2}\pi\lambda (S+2P)/\Delta E$ for the factorized and unfactorized Hamiltonians are presented in Table \ref{table:resources_methane}. 

\begin{table}[h]
\begin{center}
\begin{tabular}{ |c |c| c| c| c| c | c|} 
 \hline
  & $\lambda$ ($E_\mathrm{h}$) & Qubits & Select T count \quad & Prepare T count\quad & Total T count \quad & Relative cost \quad\\
 \hline\hline
 Unfactorized & 27.6 & 93 & $6.9\times 10 ^6$ & $7.6\times10^4$ & $2.0\times10^{11}$ & 1\\ 
 Tucker decomposition & 28.4 & 89 & $3.1\times10^7$ & $3.5\times10^5$ & $9.1\times10^{11}$ & 4.6\\ 
 CP decomposition & 8.40 & 85 & $1.1\times10^7$ & $5.6\times 10^4$ & $9.7\times10^{10}$ & 0.49\\ 
 \hline
\end{tabular}
\caption{The 1-norms, T-counts and relative cost of factorization methods for the CP and Tucker decompositions for the monodeutered methane molecule. For the CP decomposition, we used decomposition ranks corresponding to a total QPE error of 1\%. The 1-norm $\lambda$ is given in Hartrees ($E_\mathrm{h}$).}
\label{table:resources_methane}
\end{center}
\end{table}

\section{\label{sec:discussion}Discussion}

We estimated the fault-tolerant quantum computing resources, specifically the T-gate count  required for simulating a class of vibronic Hamiltonians that can be expanded as power series in internal vibrational coordinates. To this end, we applied two tensor decomposition techniques: the canonical polyadic (CP) decomposition and the Tucker decomposition. For each method, we analyzed the total T-gate and qubit requirements as functions of the Hamiltonian's structure and basis set parameters.

The effectiveness of a factorized Hamiltonian form, controlled by the decomposition rank, depends on both the error tolerance $\epsilon_\mathrm{F}$ and the specific structure of the Hamiltonian. For the H$_2$O and CH$_3$D molecules studied, we found that the CP decomposition, at a final QPE error level of 1\%, reduces the 1-norm of the vibrational and vibronic Hamiltonians. Although this decomposition results in higher T-gate counts for block encoding, it ultimately yields an overall reduction in T-gate count by more than a factor of two for CH$_3$D. For H$_2$O there is a slight increase in T-gate count when using the CP decomposition. For the Tucker decomposition, we find that the cost increases significantly (4.6 times) for the CH$_3$D simulation but we obtain a savings of 33\% for the H$_2$O molecule simulation. Evidently, the factorization method which should be used depends on the acceptable error in the simulation and the particular molecule to be simulated. The 1-10\% accuracy in the energy evaluation~\cite{motlagh2024a} corresponds to typical requirements for simulations of singlet fission for solar cell
design. We thus note that our block-encoding approach can be used with qubitization or quantum signal processing with phase estimation schemes~\cite{arrazola2024}  for calculations of vibronic energies and dynamics.


While the Tucker decomposition can also employ low-rank representations by zeroing small core tensor elements or reducing core tensor dimensionality, we found that such approximations were insufficient to accurately reconstruct the H$_2$O and CH$_3$D Hamiltonians used in this study.

We recommend careful evaluation of the decomposition rank for a given error budget when constructing approximate Hamiltonians. The CP decomposition is particularly beneficial for moderate-accuracy (1–10\%) simulations and sparsely coupled systems, which are common in material design and biochemistry. Importantly, our approach is not limited to normal coordinates: following Ref.~\cite{Rey2019}, curvilinear coordinate representations in second-quantized form can lead to more compact potential energy surface representations, where our techniques may offer additional benefits. Finally, beyond CP and Tucker decompositions, other tensor factorization methods such as tensor train\cite{oseledets2011, bigoni2016} approaches may also be applied to the simulation of vibrational and vibronic Hamiltonians.


\section{Acknowledgements}
The authors would like to thank Ákos Nagy, Konrad Deka, Michał Szczepanik, Tom Ginsberg, and the BEIT Inc. team for helpful discussions and feedback. This work was funded by the European Innovation Council accelerator grant COMFTQUA, no. 190183782. 

\bibliography{references}

\pagebreak
\widetext
\begin{center}
\textbf{\large Supplementary Information: Higher order tensor factorizations for block encoding vibrational and vibronic hamiltonians}
\end{center}
\setcounter{equation}{0}
\setcounter{figure}{0}
\setcounter{table}{0}
\setcounter{page}{1}
\setcounter{section}{0}

\makeatletter
\renewcommand{\theequation}{S\arabic{equation}}
\renewcommand{\thefigure}{S\arabic{figure}}
\renewcommand{\thesection}{S\arabic{section}}

\section{\label{si_sec:cp_decomposition}CP decomposition}
In this section we derive the CP decomposition of the vibrational and vibronic hamiltonians. 
\subsection{Vibrational hamiltonian}
We first consider vibrational hamiltonians of the form
\begin{align}
H &= H_{\mathrm{harmonic}} + \sum_{\alpha_1\alpha_2\alpha_3}E_{\alpha_1\alpha_2\alpha_3}q_{\alpha_1}q_{\alpha_2}q_{\alpha_3} + \sum_{\alpha_1\alpha_2\alpha_3\alpha_4}E_{\alpha_1\alpha_2\alpha_3\alpha_4}q_{\alpha_1}q_{\alpha_2}q_{\alpha_3}q_{\alpha_4} + \cdots \\
&= H_{\mathrm{harmonic}} + \sum_{k=3}^{L_\mathrm{v}}\left(\sum_{\alpha_1\cdots \alpha_k}E_{\alpha_1\dots \alpha_k}q_{\alpha_1}\cdots q_{\alpha_k}\right)
\label{si_eq:high_order_vibrational}
\end{align}
where $H_{\mathrm{harmonic}}=\sum_\alpha\omega_\alpha a_\alpha^\dagger a_\alpha$ has already been rewritten in terms of the bosonic creation and annihilation operators. By rewriting $q_i=(a_i+a_i^\dagger)/\sqrt{2}$ and using the bosonic commutation relations, it can be shown that $[q_i,q_j]=0$. Each tensor $E_{\alpha_1\cdots \alpha_k}$ is symmetric under arbitrary permutations of the indices $\alpha_1,\dots,\alpha_k$. 

As shown in Ref. \cite{comon2008, kolda2015}, a symmetric tensor $E_{\alpha_1\cdots \alpha_k}$ always has a symmetric CP decomposition 
\begin{equation}
    E_{\alpha_1\cdots \alpha_k}= \sum_{l=1}^{r_k}\Lambda_{kl}Q_{kl\alpha_1}\cdots Q_{kl\alpha_k}
\end{equation}
where each $Q_{kl}$ is a unit vector. We can therefore rewrite Eq.~\eqref{si_eq:high_order_vibrational} as 
\begin{align}
    H_\mathrm{v} &= H_{\mathrm{harmonic}} + \sum_{k=3}^{L_\mathrm{v}}\left(\sum_{\alpha_1\cdots \alpha_k}E_{\alpha_1\dots \alpha_k}q_{\alpha_1}\cdots q_{\alpha_k}\right)\\
    &= H_{\mathrm{harmonic}} + \sum_{k=3}^{L_\mathrm{v}}\left(\sum_{\alpha_1\cdots \alpha_k}\sum_{l=1}^{r_k}\Lambda^\mathrm{v}_{kl}Q^\mathrm{v}_{kl\alpha_1}\cdots Q^\mathrm{v}_{kl\alpha_k}q_{\alpha_1}\cdots q_{\alpha_k}\right)\\
    &= H_{\mathrm{harmonic}} + \sum_{k=3}^{L_\mathrm{v}}\left(\sum_{l=1}^{r_k}\Lambda^\mathrm{v}_{kl}\sum_{\alpha_1}Q^\mathrm{v}_{kl\alpha_1}q_{\alpha_1}\sum_{\alpha_1}Q^\mathrm{v}_{kl\alpha_1}q_{\alpha_2}\cdots\sum_{\alpha_k}Q^\mathrm{v}_{kl\alpha_k}q_{\alpha_k}\right)\\
    &= H_{\mathrm{harmonic}} + \sum_{k=3}^{L_\mathrm{v}}\left(\sum_{l=1}^{r_k}\Lambda^\mathrm{v}_{kl}(s_{kl})^k\right)\\
    &= H_{\mathrm{harmonic}} + \sum_{k=3}^{L_\mathrm{v}}\sum_{l=1}^{r_k}\Lambda^\mathrm{v}_{kl}(s_{kl})^k,
    \label{si_eq:cp_vibrational}
\end{align}
where $s_{kl} = \sum_{\alpha}Q^\mathrm{v}_{kl\alpha}q_{\alpha}$ and the superscript $\mathrm{v}$ stands for vibrational.

\subsection{Vibronic hamiltonian}
Next we consider the higher order vibronic coupling hamiltonian, which has the general form 
\begin{equation}
    H_{\mathrm{vc}} = \sum_{k=1}^{L_\mathrm{vc}}\sum_{\substack{\alpha_1\cdots\alpha_k \\ \sigma ij}}E_{\alpha_1\cdots\alpha_k\sigma ij}q_{\alpha_1}\cdots q_{\alpha_k} c_{i\sigma}^\dagger c_{j\sigma}.
    \label{si_eq:vibronic_hamiltonian}
\end{equation}
$E_{\alpha_1\cdots\alpha_k\sigma ij}$ is symmetric under arbitrary permutations of the $\alpha$ indices, as well as under swapping of the $i$ and $j$ indices, provided we use real basis functions.

Using the same decomposition of the symmetric tensor $E_{\alpha_1\cdots\alpha_k\sigma ij}$ (for each fixed $\sigma i j$), we have that
\begin{align}
    H_{\mathrm{vc}} &= \sum_{k=1}^{L_\mathrm{vc}}\sum_{\substack{\alpha_1\cdots\alpha_k \\ \sigma ij}}E_{\alpha_1\cdots\alpha_k\sigma ij}q_{\alpha_1}\cdots q_{\alpha_k} c_{i\sigma}^\dagger c_{j\sigma} \\
    &= \sum_{k=1}^{L_\mathrm{vc}}\sum_{\substack{\alpha_1\cdots\alpha_k \\ \sigma ij}}\sum_{l=1}^{r_k}\Lambda^\mathrm{vc}_{kl\sigma i j}Q^{\mathrm{vc}}_{kl\alpha_1\sigma i j}q_{\alpha_1}\cdots Q^{\mathrm{vc}}_{kl\alpha_k\sigma i j}q_{\alpha_k}c_{i\sigma}^\dagger c_{j\sigma}\\
    &= \sum_{k=1}^{L_\mathrm{vc}}\sum_{\sigma i j}\sum_{l=1}^{r_k}\Lambda^\mathrm{vc}_{kl\sigma i j}\left(\sum_{\alpha_1}Q^{\mathrm{vc}}_{kl\alpha_1\sigma i j}q_{\alpha_1}\right)\cdots \left(\sum_{\alpha_k}Q^{\mathrm{vc}}_{kl\alpha_k\sigma i j}q_{\alpha_k}\right)c_{i\sigma}^\dagger c_{j\sigma}\\
    &=\sum_{k=1}^{L_\mathrm{vc}}\sum_{\sigma i j}\sum_{l=1}^{r_k}\Lambda^\mathrm{vc}_{kl\sigma i j}\left( s_{kl\sigma i j}\right)^k c_{i\sigma}^\dagger c_{j\sigma},\label{si_eq:vibronic_decomp}
\end{align}
where $s_{kl\sigma i j}=\sum_{\alpha}Q^{\mathrm{vc}}_{kl\alpha\sigma i j}q_{\alpha}$.

\section{\label{si_sec:tucker_decomposition}Tucker decomposition}
In this section we derive the Tucker decomposition of the vibrational and vibronic hamiltonians. 

\subsection{Vibrational hamiltonian}
We start with the same general hamiltonian in Eq.~\eqref{si_eq:high_order_vibrational}. 
The Tucker decomposition, also known as the higher order singular value decomposition (HOSVD) \cite{lathauwer2000}, factorizes a tensor $E$ as 
\begin{equation}
    E_{\alpha_1\cdots \alpha_k} = \sum_{\beta_1\cdots \beta_k}\Lambda_{\beta_1\cdots \beta_k}u_{k\beta_1\alpha_1}\cdots u_{k\beta_k\alpha_k},
    \label{si_eq:tucker_decomp}
\end{equation}
where $u_j$ is a unitary matrix. The advantage of this decomposition over the CP decomposition is that there exists efficient algorithms for the exact decomposition of this form. 

If we substitute the HOSVD from Eq.~\eqref{si_eq:tucker_decomp} into the vibrational Hamiltonian, we get the decomposition
\begin{align}
    H_\mathrm{v} &= H_{\mathrm{harmonic}} + \sum_{k=3}^{L_\mathrm{v}}\left(\sum_{\alpha_1\cdots \alpha_k}E_{\alpha_1\dots \alpha_k}q_{\alpha_1}\cdots q_{\alpha_k}\right)\\
    &= H_{\mathrm{harmonic}} + \sum_{k=3}^{L_\mathrm{v}}\sum_{\alpha_1\cdots \alpha_k}\sum_{\beta_1\cdots \beta_k}\Lambda_{\beta_1\cdots \beta_k}u_{k\beta_1\alpha_1}\cdots u_{k\beta_k\alpha_k}q_{\alpha_1}\cdots q_{\alpha_k}\\
    &= H_{\mathrm{harmonic}} + \sum_{k=3}^{L_\mathrm{v}}\sum_{\beta_1\cdots \beta_k}\Lambda_{\beta_1\cdots \beta_k}\left(\sum_{\alpha}u_{k\beta_1\alpha}q_{\alpha}\right)\cdots \left(\sum_{\alpha}u_{k\beta_k\alpha}q_{\alpha}\right)\\
    &= H_{\mathrm{harmonic}} + \sum_{k=3}^{L_\mathrm{v}}\sum_{\beta_1\cdots \beta_k}\Lambda_{\beta_1\cdots \beta_k}s_{k\beta_1}\cdots s_{k\beta_k}
\end{align}
where $s_{k\beta_i} = \sum_{\alpha}u_{k\beta_i\alpha}q_{\alpha}$.

\subsection{Vibronic hamiltonian}
The Tucker decomposition of the vibronic hamiltonian has a similar form to the vibrational hamiltonian. We derive it in the following, starting from Eq.~\eqref{si_eq:vibronic_hamiltonian}.
\begin{align}
    H_\mathrm{v} &= H_{\mathrm{harmonic}} + \sum_{k=1}^{L_\mathrm{vc}}\left(\sum_{\substack{\alpha_1\cdots\alpha_k \\ \sigma ij}}E_{\alpha_1\dots \alpha_k\sigma ij}q_{\alpha_1}\cdots q_{\alpha_k}c_{i\sigma}^\dagger c_{j\sigma}\right)\\
    &= H_{\mathrm{harmonic}} + \sum_{k=1}^{L_\mathrm{vc}}\sum_{\substack{\alpha_1\cdots\alpha_k \\ \sigma ij}}\sum_{\beta_1\cdots \beta_k}\Lambda_{\beta_1\cdots \beta_k\sigma ij}u_{k\beta_1\alpha_1}\cdots u_{k\beta_k\alpha_k}q_{\alpha_1}\cdots q_{\alpha_k}c_{i\sigma}^\dagger c_{j\sigma}\\
    &= H_{\mathrm{harmonic}} + \sum_{k=1}^{L_\mathrm{vc}}\sum_{\beta_1\cdots \beta_k}\Lambda_{\beta_1\cdots \beta_k\sigma ij}\left(\sum_{\alpha}u_{k\beta_1\alpha}q_{\alpha}\right)\cdots \left(\sum_{\alpha}u_{k\beta_k\alpha}q_{\alpha}\right)c_{i\sigma}^\dagger c_{j\sigma}\\
    &= H_{\mathrm{harmonic}} + \sum_{k=1}^{L_\mathrm{vc}}\sum_{\beta_1\cdots \beta_k}\Lambda_{\beta_1\cdots \beta_k\sigma ij}s_{k\beta_1}\cdots s_{k\beta_k}c_{i\sigma}^\dagger c_{j\sigma}
\end{align}
where $s_{k\beta_i} = \sum_{\alpha}u_{k\beta_i\alpha}q_{\alpha}$.

\section{\label{si_sec:bosonic_encoding}Bosonic encoding scheme}
To calculate the norm of the above Hamiltonian, we need to choose a bosonic encoding scheme\cite{sawaya2020}. Here we choose a unary encoding, since it requires fewer Pauli operators to represent the position operator versus a binary/gray encoding scheme.

For the unary encoding, the bosonic states are encoded as 
\begin{equation}
    \begin{cases}
        \ket{0} \leftrightarrow \ket{1 0 0\cdots 0},\\
        \ket{1} \leftrightarrow \ket{0 1 0\cdots 0},\\
        \ket{2} \leftrightarrow \ket{0 0 1\cdots 0},\\
        \vdots\\
        \ket{d} \leftrightarrow \ket{0 0 \cdots 0 1}\\
    \end{cases}
\end{equation}
and requires $d+1$ qubits if we truncate our Fock space to include up to $d$ phonons. Writing $\sigma_+=(X-iY)/2$ we can then represent the raising operator as
\begin{equation}
    a^\dagger = \sum_{i=0}^{d-1}\sqrt{i+1}\sigma_-^i\sigma_+^{i+1}
\end{equation}
and the position operator as 
\begin{equation}
    q=\frac{a + a^\dagger}{\sqrt{2}} = \frac{1}{2\sqrt{2}}\sum_{i=0}^{d-1}\sqrt{i+1}\left(X_iX_{i+1} + Y_iY_{i+1}\right).
\end{equation}
The number operator $a^\dagger a$ is represented by the qubit operator
\begin{equation}
    a^\dagger a = \sum_{i=1}^di\sigma_+^i\sigma_-^i.
\end{equation}

\section{\label{si_sec:vibronic_hamiltonian}Vibronic Hamiltonian coefficients}
For the purpose of demonstrating the tensor decompositions, we model the vibronic coupling Hamiltonian as a perturbation of the vibrational Hamiltonian. In particular, for a given non-zero vibrational coefficient $E_{\alpha_1\cdots\alpha_k}$, we construct a corresponding vibronic coefficient as
\begin{equation}
    E_{\alpha_1\cdots\alpha_k\sigma i j}=(E_{\alpha_1\cdots\alpha_k} + \eta_{\alpha_1\cdots\alpha_k\sigma i j}) \frac{2 ^ {-k}}{\max(1, |i-j|)}
\end{equation}
where $\eta_{\alpha_1\cdots\alpha_k\sigma i j}$ is drawn from a uniform distribution centered at 0 and with width 1/20. We emphasize that this is not intended to precisely model the vibronic interaction Hamiltonian but to provide a physical model to benchmark the different tensor decomposition methods. 

\section{\label{si_sec:block_encoding} Block encoding circuits}
Here we present the circuits which are used to block encode the hamiltonian. We use the notation of Reference \cite{burg2021}. The three main circuit primitives are the circuits for block encoding a hamiltonian represented as a linear combination of unitaries, circuits for adding block encodings, and circuits for multiplying block encodings. 
The circuit for block encoding a hamiltonian $H=\sum_xa_xU_x$, $\lambda=\sum_x|a_x|^2$ is represented as

\tikzset{
     hex/.style={
         thickness, 
         filling, 
             transparent.
         shape=circle, 
         draw=black, 
         inner sep=2pt
}}

\begin{equation}
\begin{quantikz}
    \lstick{$\ket{0}_a$} &  \gate[2]{\mathrm{B}[H/\lambda]}  & \\
    \lstick{$\ket{\psi}_s$} &  & 
\end{quantikz}=
\begin{quantikz}
    \lstick{$\ket{0}_a$} & \gate{\mathrm{Prepare}(a)} & \phase{x}\wire[d]{q} & \gate{\mathrm{Prepare}(a)^\dagger} & \\
    \lstick{$\ket{\psi}_s$} & &\gate{\{U_x\}}& & 
\end{quantikz}
\label{si_eq:lcu}
\end{equation}

\noindent where $\mathrm{Prepare}(a)\ket{0}_a=\sum_xa_x\ket{x}_a$.
The middle gate in the second circuit represents the multiplexed unitary which maps $\ket{x}_a\ket{\psi}_s\mapsto\ket{x}_aU_x\ket{\psi}_s$
Given multiple hamiltonians $H_i$ each written as a sum of unitaries, we can block encode linear combinations of the hamiltonians $\sum_ia_iH_i$ with the circuit 

\begin{equation}
\begin{quantikz}
    \lstick{$\ket{0}_{a_1}$} &  \gate[3]{\mathrm{B}[\sum_ia_iH_i]}  & \\
    \lstick{$\ket{0}_{a_2}$} & &\\
    \lstick{$\ket{\psi}_s$} &  & 
\end{quantikz} = \begin{quantikz}
    \lstick{$\ket{0}_{a_1}$} & \gate{\mathrm{Prepare}(a)} & \phase{i}\wire[d]{q} & \gate{\mathrm{Prepare}(a)^\dagger} &  \\
    \lstick{$\ket{0}_{a_2}$} & & \gate[2]{\{B[H_i]\}}  & &\\
    \lstick{$\ket{\psi}_s$} & & & &
\end{quantikz}
\label{si_eq:lcb}
\end{equation}
Here, ancilla $a_2$ is used in the block encodings of the individual hamiltonians. 

The final circuit primitive is the block encoding of the multiplication of hamiltonians $B[H_1\cdots H_n]$ and is given by the circuit

\begin{equation}
\begin{quantikz}
    \lstick{$\ket{0}_{a_1}$} &  \gate{X}  & \targ{} & & & \ldots&&& \gate{X} &\\
    \lstick{$\ket{0}_{a_2}$} & \gate{X} & & &\targ{} & \ldots &&& \gate{X} &\\
    \lstick{\vdots} \\
    \lstick{$\ket{0}_{a_n}$} & \gate{X} &  &&&&&\targ{}& \gate{X} &\\
    \lstick{$\ket{0}_{a_{n+1}}$} & \gate[2]{B[H_1]} & \ctrl[open]{-4} & \gate[2]{B[H_2]} & \ctrl[open]{-3} &\ldots & \gate[2]{B[H_n]} &\ctrl[open]{-1}&&\\[0.5cm]
    \lstick{$\ket{\psi}_s$} &  &  &&& &&&&
\end{quantikz}
\label{si_eq:multiplying_block_encodings}
\end{equation}
Here, ancilla $a_{n+1}$ is used in the block encodings of the individual hamiltonians and the open control on multiple qubits is a multi-controlled operator.

These block encoding primitives are used as follows. First, we write the linear combination of position operators as a sum of unitaries using the unary bosonic encoding scheme of Section \ref{si_sec:bosonic_encoding}. This linear combination of unitaries can be block encoded using the circuit in Eq.~\ref{si_eq:lcu}. To get higher powers of these linear combinations, we use the multiplying circuit in Eq.~\ref{si_eq:multiplying_block_encodings}. Finally, to sum all the terms in the hamiltonian we use the linear combination of block encodings on Eq.~\ref{si_eq:lcb}.

\section{\label{si_sec:asymptotic_gate_count} Asymptotic T gate counts}
In this section we derive the gate counts for a single block encoding in terms of the parameters of hamiltonian: the number of electron orbitals $N$, the number of vibrational modes $M$, the ultraviolet cutoff (number of phonon modes retained) $d$, the vibrational cutoff $L_\mathrm{v}$, and the vibronic cutoff $L_{\mathrm{vc}}$. The CP decomposition additionally has a rank parameter $r$ for determining the level of approximation. For simplicity, we assume that the parameters $d$ and $r$ are constant for each order in the taylor series expansion. The block encoding cost can be decomposed as $S+2P$, where $S$ is the cost of Select and $P$ is the cost of Prepare. We note that since we are using recursive block encodings, the inner block encodings have their own Select and Prepare costs. We combine all the Selects from all inner block encodings and all Prepares from all inner block encodings to simplify the analysis. Counting in this way does not affect the gate count.

We first derive gate counts for the CP decomposition, starting with the Select method. We begin with the quadratic part of the vibrational hamiltonian $\sum_\alpha\omega_\alpha a_\alpha a_\alpha^\dagger$, which is not decomposed. Using a unary decomposition, each term $a_\alpha a_\alpha^\dagger$ can be written as a sum of $d$ unitaries. As there are $M$ vibrational modes, we can implement the Select routine of the quadratic vibrational hamiltonian using unary iteration\cite{babbush2018} with $O(Md)$ T gates. 

For the vibrational terms of higher order, we focus on an individual term $\sum_{l=1}^{r}\Lambda^\mathrm{v}_{jl}(s_{jl})^j$
where $s_{jl} = \sum_{\alpha}Q^\mathrm{v}_{jl\alpha}q_{\alpha}$. Each term $s_{jl}$ can be written as a sum of $2Md$ unitaries, and so a block encoding requires $O(Md)$ T gates. The multiplication of $j$ block encodings requires $j$ $\mathrm{C}_nX$ gates, where the number of controls $n$ is the size of the ancilla register required for the block encodings. Here the ancilla register contains $\log(2Md)$ qubits which must be controlled; the multicontrolled gates therefore require $O(j\log(2Md))$ T gates\cite{gidney2015}. A single block encoding of the product $(s_{jl})^j$ therefore requires $O(jMd)$ T gates. 

The total cost of implementing the Select operations for the vibrational hamiltonians is then given by
\begin{equation}
O\left(r\sum_{j=3}^{L_\mathrm{v}}1\right)+\sum_{j=3}^{L_\mathrm{v}}O(rjMd)=O\left(rL_\mathrm{v}\right)+O\left(rMdL_\mathrm{v}^2\right)=O\left(rMdL_\mathrm{v}^2\right),
\end{equation}
where the first term on the left is the T count of adding the individual block encodings together. 
The gate count of the vibronic hamiltonian differs only in the final step, where we need to include the electron orbitals, so the final T gate count for the Select subroutine in the CP decomposition is
\begin{equation}
O\left(rMd\left(L_\mathrm{v}^2+L_\mathrm{vc}^2N^2\right)\right).
\end{equation}

We next derive an upper bound for the T gate count for the Prepare operation for the CP decomposition. Throughout, we use the Select-Swap method\cite{low2024} with $\lambda=\sqrt{N}$ for preparing $N$ elements with precision $\epsilon$, which results in a T count of $O(\sqrt{N}\log\frac{N}{\epsilon})$.

First, we count the total number of Prepare operations that will be needed, as this will be required to determine the accuracy required for each Prepare. For the quadratic vibrational term, we need a single Prepare operation with $O(Md)$ terms.
For the vibrational and vibronic hamiltonians, we need a Prepare operation each time we block encode a linear combination $s_{jl} = \sum_{\alpha}Q^\mathrm{v}_{jl\alpha}q_{\alpha}$. There are a total of
\begin{equation}
\sum_{j=3}^{L_\mathrm{v}}rj+\sum_{j=1}^{L_\mathrm{vc}}rj=O(r(L_\mathrm{v}^2+L_\mathrm{vc}^2))
\end{equation}
of these sub-Prepare operations as well as the two Prepare operations needed in the outer block encodings. We therefore have a total number of Prepare operations bounded by $O(r(L_\mathrm{v}^2+L_\mathrm{vc}^2))$. If we wish to have a total allowable error of $\epsilon_{\mathrm{Prep}}$, we need to implement each sub-Prepare operation with precision
\begin{equation}
    \epsilon'=O\left(\frac{\epsilon_{\mathrm{Prep}}}{r(L_\mathrm{v}^2+L_\mathrm{vc}^2)}\right). 
\end{equation}
We now count the gate cost of implementing all the sub-Prepare routines. The quadratic term in the vibrational hamiltonian has $O(Md)$ terms and incurs a T cost of $O(\sqrt{Md}\log\frac{Md}{\epsilon'})$. 

Next we find the T count of the higher order vibrational hamiltonian. Each term $s_{jl}$ requires a Prepare operation with $O(Md)$ terms and costs $O(\sqrt{Md}\log\frac{Md}{\epsilon'})$ T gates. 
The product $(s_{jl})^j$ therefore requires $O(j\sqrt{Md}\log\frac{Md}{\epsilon'})$ T gates. The T count from all the products that need to be implemented is therefore
\begin{equation}
\sum_{j=3}^{L_\mathrm{v}}O\left(rj\sqrt{Md}\log\frac{Md}{\epsilon'}\right)=O\left(rL_\mathrm{v}^2\sqrt{Md}\log\frac{Md}{\epsilon'}\right).
\end{equation}
There are $O(rL_\mathrm{v})$ terms that need to be added together in the outer block encoding, which incurs a T cost \begin{equation}O\left(\sqrt{rL_\mathrm{v}}\log\frac{rL_\mathrm{v}}{\epsilon'}\right)
\end{equation}
so the total T count for all the Prepare subroutines in the vibrational hamiltonian is given by
\begin{equation}
    O\left(rL_\mathrm{v}^2\sqrt{Md}\log\frac{Md}{\epsilon'}+\sqrt{rL_\mathrm{v}}\log\frac{rL_\mathrm{v}}{\epsilon'}\right)
\end{equation}
Similarly to the case for Select, the cost of the vibronic term differs in the total number of terms that need to be summed, so the T count for the vibronic hamiltonian is 
\begin{equation}
    O\left(rL_\mathrm{v}^2\sqrt{Md}\log\frac{Md}{\epsilon'}+N\sqrt{rL_\mathrm{vc}}\log\frac{rL_\mathrm{vc}N^2}{\epsilon'}\right)
\end{equation}
Adding all the T counts of the Select and Prepare components of the various hamiltonians results in a final T count of 
\begin{align}
O\left(rMd\left(L_\mathrm{v}^2+L_\mathrm{vc}^2N^2\right) + 
rL_\mathrm{v}^2\sqrt{Md}\log\frac{Md}{\epsilon'}+\sqrt{rL_\mathrm{v}}\log\frac{rL_\mathrm{v}}{\epsilon'} +
rL_\mathrm{vc}^2\sqrt{Md}\log\frac{Md}{\epsilon'}+N\sqrt{rL_\mathrm{vc}}\log\frac{rL_\mathrm{vc}N^2}{\epsilon'}\right)
\end{align}
where 
\begin{equation}
    \epsilon'=\frac{\epsilon_\mathrm{Prep}}{r(L_\mathrm{v}^2+L_\mathrm{vc}^2)}
\end{equation}

The analysis for the Tucker decomposition is similar. The difference is that the rank $r$ which was held fixed in the CP decomposition analysis now corresponds to the number of non-zero terms in the Tucker decomposition, which is at most $M^j$, where $j$ is the order of the taylor series expansion of the vibrational or vibronic interaction hamiltonian. 

The cost of the Select for the vibrational hamiltonian then becomes 
\begin{equation}
O\left(\sum_{j=3}^{L_\mathrm{v}}M^j\right)+\sum_{j=3}^{L_\mathrm{v}}O(M^jjMd)=O\left(M^{L_\mathrm{v}}\right)+O\left(dL_\mathrm{v}M^{L_\mathrm{v}+1}\right)=O\left(dL_\mathrm{v}M^{L_\mathrm{v}+1}\right),
\end{equation}
and the Select cost for the vibronic hamiltonian is 
\begin{equation}
O\left(N^2dL_\mathrm{vc}M^{L_\mathrm{vc}+1}\right).
\end{equation}

A similar analysis for the Prepare cost can be carried out by replacing the constant rank $r$ with $M^j$ for each order $j$. There are now a total of $L_\mathrm{v}M^{_\mathrm{v}}+L_\mathrm{vc}M^{_\mathrm{vc}}+2$ Prepare operations that are needed, so we implement each Prepare with a precision of
\begin{equation}
    \epsilon'=O\left(\frac{\epsilon_\mathrm{Prep}}{L_\mathrm{v}M^{L_\mathrm{v}}+L_\mathrm{vc}M^{L_\mathrm{vc}}}\right).
\end{equation}
The resulting Prepare T count for the vibrational hamiltonian is
\begin{equation}
O\left(L_\mathrm{v}M^{L_\mathrm{v}+1/2}\sqrt{d}\log\frac{Md}{\epsilon'} + M^{L_\mathrm{v}/2}\log\frac{M^{L_\mathrm{v}}}{\epsilon'}\right)
\end{equation}
and 
\begin{equation}
O\left(L_\mathrm{vc}M^{L_\mathrm{vc}+1/2}\sqrt{d}\log\frac{Md}{\epsilon'} + NM^{L_\mathrm{vc}/2}\log\frac{M^{L_\mathrm{vc}}N^2}{\epsilon'}\right)
\end{equation}
for the vibronic hamiltonian.
The total T count is the sum of all three components above and is given by
\begin{align}
    O\left(dL_\mathrm{v}M^{L_\mathrm{v}+1} + N^2dL_\mathrm{vc}M^{L_\mathrm{vc}+1} + L_\mathrm{v}M^{L_\mathrm{v}+1/2}\sqrt{d}\log\frac{Md}{\epsilon'} \right. &+ \\  M^{L_\mathrm{v}/2}\log\frac{M^{L_\mathrm{v}}}{\epsilon'} + L_\mathrm{vc}M^{L_\mathrm{vc}+1/2}\sqrt{d}\log\frac{Md}{\epsilon'} &+ \left.NM^{L_\mathrm{vc}/2}\log\frac{M^{L_\mathrm{vc}}N^2}{\epsilon'}\right)
\end{align}

We note that there is a final addition of block encodings required to add the block encodings of $H_\mathrm{harmonic}$, $H_\mathrm{v}$ and $H_\mathrm{vc}$ (as well as the omitted electronic hamiltonians). However, this is a constant overhead which we omit for the purposes of asymptotic analysis.

\section{\label{si_sec:error_analysis} Error analysis}
There error analysis presented here follows Reference \cite{babbush2018}. When applying phase estimation to estimate the eigenvalues of a hamiltonian $H_s$ with norm $\lambda$, we can use qubitized quantum walks \cite{low2019} to implement the unitary operator $\exp(i\arccos(H_s/\lambda))$ exactly when given access to qubitization oracles $B$ such that $\left(\bra{0}_a\otimes \identity_s\right)B_{as}\left(\ket{0}_a\otimes \identity_s\right)=H_s/\lambda$ where the subscripts indicate which register an operator acts on \cite{berry2018}.

\subsection{Accuracy required for estimating eigenvalues of $\exp(iW(H/\lambda))$}
Suppose $W:[-1,1]\rightarrow[-1,1]$ is  invertible and we can implement $\exp(iW(H/\lambda)2^j)$ exactly for integers $j\geq0$. The outcome of a phase estimation circuit with precision $\delta$ will be a number $x=W(E_k/\lambda)+\delta_0$, where $E_k$ is an eigenvalue of $H$ and $|\delta_0|\leq\delta$. Since $W$ is invertible on $[-1,1]$ and $|E_k/\lambda|<1$ by construction, we can write
\begin{align}
    \frac{E_k}{\lambda} &= W^{-1}(x-\delta_0)\\
    &= W^{-1}(x)-(W^{-1})'(x)\delta_0 + O(\delta_0^2)\\
    &=W^{-1}(x)-\frac{\delta_0}{W'(x)} + O(\delta_0^2)\\
    &\implies |E_k - \lambda W^{-1}(x)| = \lambda\left|\frac{\delta_0}{W'(x)}\right| +  O(\lambda\delta_0^2).
\end{align}
Since $W^{-1}(x)$ is the value we calculate from the outcome of phase estimation on $\exp(i\arccos(H/\lambda))$, we see that to obtain $E_k$ to precision $\epsilon$ we need to perform phase estimation of $\arccos(H/\lambda)$ to precision $\epsilon$ such that $\lambda\left|\delta_0/||W'(x)||\right|<\epsilon$ for all $x\in[-1,1]$. In other words, we need to perform phase estimation of $\exp(i\arccos(H/\lambda))$ to precision \begin{equation}\frac{\epsilon}{\lambda}\left[\inf_{x\in[-1,1]}||W'(x)||\right].\end{equation}
For $W(x)=\arccos(x)$, $\inf_{x\in[-1,1]}||W'(x)||=1$.

Let $w=\inf_{x\in[-1,1]}||W'(x)||$. For a given precision $\epsilon w /\lambda$, we can write $\epsilon w / \lambda<2^{-b}$ for some integer $b$. To obtain an estimate of the phase of the unitary $\exp(i\arccos(H/\lambda))$ to precision $\epsilon w /\lambda$, we therefore need to implement phase estimation using $b$ bits of precision which requires implementing controlled versions of $\exp(i\arccos(H/\lambda)2^j)$ for $j=0,\dots,b-1$. Since we require one qubitization oracle call to implement $\exp(i\arccos(H/\lambda))$, implementing the entire circuit requires $2^b - 1 \leq \floor{\lambda/(\epsilon w)}-1$ oracle calls (we also need multicontrolled $Z$ gates to implement the quantum walk reflection step; however, this cost is negligible compared to the cost of block encoding). 

The gate cost of estimating an eigenvalue $E_k$ of $H$ to precision $\epsilon$ is therefore the gate cost of the phase estimation procedure for $\exp(i\arccos(H/\lambda))$ to precision $\epsilon w / \lambda$. This cost is given by $O\left(\frac{B\lambda}{\epsilon w}\right)$, where $B$ is the gate cost of implementing the qubitization oracle. For the function $W(x)=\arccos(x)$, $w=1$, and the total gate cost of phase estimation is given by 
\begin{equation}
    O\left(\frac{B\lambda}{\epsilon}\right).
\end{equation}

In the previous analysis, we have assumed that our block encoding procedure block encodes the hamiltonian exactly and that we implement the final inverse QFT exactly. We can get a more precise estimate of the gate count by taking into account the error of the block encoding procedure (which is due to the errors incurred by the Prepare operation), the error in the low rank approximation, and the error in implementing an approximate QFT. 

Following Reference\cite{babbush2018}, we break up the estimated phase into a sum of 5 contributions,
\begin{equation}
\phi_{\mathrm{est}}=\phi+\epsilon_{\mathrm{Prep}}+\epsilon_{\mathrm{F}}+\epsilon_{\mathrm{QFT}}+\phi_{\mathrm{true}}
\end{equation}
where $\phi$ is a random variable with $\expect{\phi}=0$ and Holevo variance $\variance{\phi}=\tan^2(\pi/(2^{m+1}+1))\approx \pi/2^{m+1}$ and describes the optimal precision phase estimation that can be achieved with $m$ ancillary qubits. $\epsilon_{\mathrm{Prep}}$ is the error incurred by the Prepare subroutine due to the Clifford+T decomposition of the required controlled rotations in the Select-Swap procedure used in Prepare. $\epsilon_{\mathrm{QFT}}$ is the error incurred due to the Clifford+T decomposition of the controlled rotations in the inverse QFT, as well as approximation errors when using an approximate QFT. $\epsilon_{\mathrm{F}}$ is the error in factorizing the hamiltonian, $\epsilon_{\mathrm{F}}=||H-H_\mathrm{F}||_2$, where $H$ is the original hamiltonian and $H_\mathrm{F}$ is the factorized hamiltonian. We measure the error in phase as the RMS difference between the estimated phase and the true phase,
\begin{align}
    \Delta\phi &= \sqrt{\expect{(\phi_{\mathrm{est}} -\phi_{\mathrm{true}})^2}} \\
    &\approx\sqrt{\left(\frac{\pi}{2^{m+1}}\right)^2 + (\epsilon_{\mathrm{Prep}} +\epsilon_{\mathrm{F}}+\epsilon_{\mathrm{QFT}} )^2}.
\end{align}

As we are estimating the phase of $\exp(i\arccos(H/\lambda))$, the error $\delta\phi$ corresponds to a resulting error in the energy measurement as 
\begin{equation}
    \Delta E = \lambda\cos(\Delta\phi)\leq \lambda\Delta\phi\approx\lambda\sqrt{\left(\frac{\pi}{2^{m+1}}\right)^2 + (\epsilon_{\mathrm{Prep}} +\epsilon_{\mathrm{F}}+\epsilon_{\mathrm{QFT}} )^2}.
\end{equation}

To find the error in energy as a function of the number of bits of precision, we can give equal weight to both terms in the square root, i.e. we set
\begin{equation}
    \left(\frac{\pi}{2^{m+1}}\right)^2=\frac{1}{2}\left(\frac{\Delta E}{\lambda}\right)^2,\quad (\epsilon_{\mathrm{Prep}} +\epsilon_{\mathrm{F}}+\epsilon_{\mathrm{QFT}} )^2 = \frac{1}{2}\left(\frac{\Delta E}{\lambda}\right)^2.
    \label{si_eq:bits_error}
\end{equation}
In the first equality, we solve for the number of bits $m$, and find that 
\begin{equation}
    m=\ceil{\log\frac{\sqrt{2}\pi\lambda}{2\Delta E}}<\log\frac{\sqrt{2}\pi\lambda}{\Delta E}.
\end{equation}
Assigning equal weight to all three terms in the second equality of Eq.~\eqref{si_eq:bits_error}, we can choose the approximation errors as
\begin{equation}
    \epsilon_{\mathrm{Prep}} \leq \frac{1}{3\sqrt{2}}\frac{\Delta E}{\lambda}, \quad \epsilon_{\mathrm{F}} \leq \frac{1}{3\sqrt{2}}\frac{\Delta E}{\lambda}, \quad \epsilon_{\mathrm{QFT}} \leq \frac{1}{3\sqrt{2}}\frac{\Delta E}{\lambda}.
\end{equation}
As pointed out in Reference\cite{babbush2018}, this equal subdivision of error is not necessarily optimal, as reducing the error in phase estimation (increasing $m$), requires exponentially more gates than reducing the error in state preparation and QFT. 

If we ignore the cost of preparing the initial state of the ancilla register and the cost of QFT (they are both linear in $m$, while the cost of phase estimation is exponential in $m$), and assume that the Prepare subroutine has a gate cost $P$, which will depend on $\epsilon_{\mathrm{Prep}}$, and that the Select subroutine has a gate cost of $S$, then the total gate count of the phase estimation procedure is bounded above by
\begin{equation}
    \frac{\sqrt{2}\pi\lambda(S+2P)}{\Delta E}.
\end{equation}

\end{document}